\documentstyle[emulateapj,epsf]{article}

\lefthead{Wills et al.}
\righthead{A Radio Loud BAL QSO}

\begin{document}

\title{PKS\,1004+13: A High Inclination, Highly Absorbed Radio-Loud QSO --
The First Radio-Loud BAL QSO at Low Redshift?\altaffilmark{1}}

\author{Beverley J. Wills\altaffilmark{2},
W. N. Brandt\altaffilmark{3} \&
A. Laor\altaffilmark{4}}

\altaffiltext{1}{Based partly on observations by the International Ultraviolet Explorer
satellite, collected at the Villafranca Satellite Tracking Station of the European Space Agency.}
\altaffiltext{2}{McDonald Observatory \& Astronomy Department, University of
Texas at Austin, TX 78712, USA; bev@astro.as.utexas.edu}
\altaffiltext{3}{Department of Astronomy \& Astrophysics, 525 Davey Laboratory,
Pennsylvania State University, University Park, PA 16802; niel@astro.psu.edu}
\altaffiltext{4}{Department of Physics, Technion, Israel Institute of
Technology, Haifa 32000, Israel; laor@physics.technion.ac.il}

\authoremail{bev@astro.as.utexas.edu}

\begin{abstract}

The existence of BAL outflows in only radio-quiet QSOs was thought to be 
an important clue to mass ejection and the radio-loud -- radio-quiet dichotomy.
Recently a few radio-loud BAL QSOs have been discovered at high redshift. 
We present evidence that PKS\,1004+13 is a radio-loud BAL QSO.  It would be
the first known at low-redshift ($z_{\rm em} = 0.24$), and one of the most radio luminous.
For PKS 1004+13, there appear to be broad absorption troughs of O\,VI, N\,V, Si\,IV, and C\,IV,
indicating high-ionization outflows up to $\sim$10\,000 km s$^{-1}$.  There are also two
strong, broad ($\sim 550$\ km s$^{-1}$), high ionization, associated absorption systems 
that show partial covering of the continuum source.
The strong UV absorption we have detected suggests that the extreme soft-X-ray weakness
of PKS\,1004+13 is primarily the result of absorption.
The large radio-lobe dominance indicates BAL and associated gas at high inclinations to the
central engine axis, perhaps in a line-of-sight that passes through an accretion disk wind.

\end{abstract}


\placefigure{fig1}
\placefigure{fig2}
\placefigure{3}
\placetable{tbl-1}

\section{Introduction}

About 11\% of radio-quiet QSOs show strong broad absorption troughs (BALs) displaced to the blue
of the corresponding broad emission lines of O\,VI\,$\lambda$1034, Ly$\alpha$, N\,V\,$\lambda$1240,
Si\,IV\,$\lambda$1400, and C\,IV\,$\lambda$1549, indicating high-ionization outflows up to
0.1c -- 0.2c.  Except for the absorption, the ultraviolet spectra of BAL QSOs are very similar 
to non-BAL QSOs.  This,
together with limits on resonance photons scattered into the absorption troughs, and partial
covering deduced from spectropolarimetry and absorption-line ratios, suggests that BAL outflows
are common to all (radio-quiet) QSOs, but are recognized only when BAL material intercepts our
line-of-sight.  BALs appeared to be a unique property of radio-quiet QSOs.
Seemingly this was an important clue to central engine physics: QSOs appear to produce either
a highly collimated, powerful, relativistic jet, or a sub-relativistic uncollimated wind that
accelerates BAL clouds, but not both (Stocke et al. 1992, Weymann 1997).  Recently, with the
advent of systematic identification of faint (mJy) radio sources at 1.4GHz, several radio-loud
BAL QSOs have been discovered, so the incidence of BALs no longer changes at the conventional limit
separating radio-quiet and radio-loud QSOs (Brotherton et al. 1998a; Becker et al. 1999).
Nevertheless, BAL properties probably still provide important clues to the production of powerful
radio jets (Weymann 1997), and the axis defined by powerful radio jets may be used to relate the 
geometry of BAL regions to the symmetry axis of the central engine.

Another clue to the physics of mass ejection in BAL QSOs, and therefore perhaps in all QSOs, is
the extreme weakness of their soft X-ray fluxes (0.2keV--2keV; Green et al. 1995; Green \& Mathur
1996); even their more penetrating harder X-rays are extremely weak (2keV--10keV, Gallagher
et al. 1999).  In order to understand the nature of soft X-ray weak (SXW) QSOs,
that is, those having $\alpha_{ox} > 2$ (3000\AA\ to 2keV),
we have investigated the UV spectra of all PG QSOs with archival IUE and HST ultraviolet
spectra (Brandt et al. 1999).
PKS\,1004+13 is a member of that sample, and is also the only radio-loud, X-ray-weak member.

PKS\,1004+13 (4C\,13.41, PG\,1004+130) is a V=15.2$^m$, QSO of low redshift, $z = 0.240$, thus
$M_{\rm B}=-25.6$\footnote{H$_0 = 50$\,km s$^{-1}$ Mpc$^{-1}$, and q$_0 = 0$.}.  It is
a classic powerful lobe-dominant radio source -- 1.2\,Jy at 1.4GHz, or log\,L$_{1.4\,{\rm GHz}} \sim 
33.6$\,erg s$^{-1}$ Hz$^{-1}$, with radio-loudness, log$R^* =2.32$ (as defined by Sramek \&
Weedman 1980).
Elvis \& Fabbiano (1984) noted its exceptionally weak X-ray emission, deriving
a spectral index $>$2.01 between 2500\AA\ and 2\,keV.
They favored an explanation in terms of an exceptionally strong optical-ultraviolet Big Blue Bump,
rather than absorption of the X-rays.  They also suggested that a relatively weak EUV ionizing
continuum could result in weak BLR emission.
Here we present and discuss evidence that PKS 1004+130 is not only a powerful radio source, but also
a BAL QSO. 

\section{Observations \& Measurements}

We have retrieved and combined the NEWSIPS-calibrated IUE data from the short-wavelength
camera, including our own IUE observation
(SWP04942, SWP10850, SWP16824, SWP27517 with observation dates UT 1979 April 15, 1980 Dec. 1,
1982 April 24, 1986 Jan. 12 respectively), weighting these according to the square of the
signal-to-noise ratio.  All observations were at low dispersion in the large aperture,
giving a resolution of $\sim$1100 km s$^{-1}$\ from 1170\AA\ to 1980\AA. 
Wavelength scales were checked using the geocoronal Ly$\alpha$ line.
We have applied an approximate correction for IUE artifacts and the wings of the geocoronal
Ly$\alpha$\ line, by subtracting the SWP artifact spectrum appropriate to point source
spectral extraction, as presented by Crenshaw, Bruegman, \& Norman (1990).

By good fortune, the line-of-sight to PKS 1004+13 happens to pass only 34$\arcmin$
from the center of the Leo I dwarf spheroidal galaxy, so Bowen et al. (1997) used this QSO
as a background probe of Leo I's halo.  They obtained high-resolution (0.14\AA\,pixel$^{-1}$)
HST GHRS spectra of PKS 1004+13 from 1276\AA\ to 1562\AA.
We have retrieved the HST archival spectra, applying the same wavelength
shift as Bowen et al. to convert to a heliocentric velocity scale.

Both the shape and flux density level of the GHRS spectrum agree well with the IUE spectrum,
suggesting the correctness of the flux-density calibration, despite the possibility of a 1500\AA\
(observed wavelength) bump artifact in the IUE SWP data (Kinney et al. 1991).

\section{Results}

The HST and coadded IUE spectrum for PKS\,1004+13 is shown as the lower spectrum in Figure 1.
Beneath is a standard deviation spectrum derived from the scatter among the individual IUE 
spectra.  The systemic redshift 
$z_{\rm em} =$0.2401 is derived from our and Stockton \& MacKenty's (1987) measured wavelength
of the narrow [O\,III]\,$\lambda$5007
emission line.  Based on this, above the spectrum are shown the expected wavelengths of 
redshifted broad emission lines.  As noted by Kinney et al. (1991): ``Based on this spectrum 
alone, 1004+130 would not be classified as a QSO because of the very weak Ly$\alpha$ emission
line.'' Even the
expected strong C\,IV\,$\lambda$1549 emission is not clearly present.  To demonstrate the unusual
weakness of the broad Ly$\alpha$ emission, we note that its rest equivalent width EW = 17\AA\ is the
smallest of 19 low-redshift, radio-loud, lobe-dominant QSOs from the HST archives.
The mean for this sample is 140\AA.  The next smallest
EW(Ly$\alpha$) is 56\AA\ for 3C\,288.1, for which strong associated absorption is clearly
present (Wills et al. 1995).
Below, we compare the UV spectrum of PKS\,1004+13 with that of a typical QSO.

Kinney et al. (1991)
noted unresolved absorption features at  1534\AA\ and 1919\AA\ that they attributed to N\,V and
C\,IV doublets at a redshift of 0.24.  The high-resolution, high signal-to-noise-ratio GHRS
spectrum shows that
the components of the N\,V\,$\lambda\lambda$1238,1242 doublet are further split and that
there is corresponding strong absorption in O\,VI\,$\lambda\lambda$1032,1038
(Fig. 2a, b).  These two absorption systems are also seen, but less clearly, in Ly$\alpha$ 
(Fig. 2c).  All
these features are also marked beneath the spectrum of Fig. 1.  These systems have
heliocentric absorption redshifts of $z_a$ = 0.2364 and 0.2387, and are intrinsically broad --
respectively $\sim$600 km s$^{-1}$\ and 500 km s$^{-1}$\ total width, 
representing outflows of 50$\pm$50 km s$^{-1}$\ (no outflow) to 1200 km s$^{-1}$.
It is therefore not surprising that the corresponding high-ionization emission lines should
be suppressed over this velocity range.
There is no sign of low-ionization absorption at these redshifts in the broad Mg\,II$\lambda$2798
emission line (Antonucci, Wills, unpublished).
Both O\,VI doublets appear optically thick (${\tau}_{1034{\rm\AA}} \sim 6$), with absorbing
gas covering only
70\% to 80\% of the continuum source (using the method of Arav et al. 1999).  Thus the absorption
is intrinsic to the QSO.
The result for the N\,V doublets are consistent, but with larger uncertainties.

We also note the probable
existence of broad absorption troughs at even greater outflow velocities -- up to $\sim$10\,000
km s$^{-1}$. The wavelength range corresponding to
0 to $-$10\,000 km s$^{-1}$\ is indicated by shading in Figure 1, for O\,VI\,$\lambda$1034,
Ly$\alpha$, N\,V\,$\lambda$1240, Si\,IV\,$\lambda$1397, and C\,IV\,$\lambda$1549.
The C\,IV associated and broader absorption are visible on the 
IUE two-dimensional images for the 3 long-exposure spectra.
These BAL-like troughs suggest a reason why the Ly$\alpha$-N\,V\,$\lambda$1240 broad emission
is so weak: it has been suppressed largely by N\,V absorption.  The Ly$\alpha$\ BAL would
appear to be much weaker than for the higher ionization species.
Above the PKS 1004+13 spectrum, we show for comparison a spectrum of the well-known BAL QSO,
PG\,0946+301 (Korista \& Arav 1997), redshifted to align the broad absorption features as
indicated by the shaded strips.
Note the suppression of broad Ly$\alpha$ emission by the N\,V BAL for PG\,0946+301.
Higher quality UV spectroscopy would confirm the reality of BALs in PKS 1004+13.

One way in principle to distinguish absorption from emission, and isolate the absorption
spectrum of PKS\,1004+13, is to divide by a `normal' QSO spectrum.  We and others (e.g., Wills et al.
1999) have shown that the greatest spectrum-to-spectrum differences can be described by
luminosity relationships (the Baldwin effect), dependence on Boroson and Green's (1992)
optical Principal Component 1 (PC1), and radio core-dominance (Baker \& Hunstead 1995,
Vestergaard 1998).
Therefore we choose to divide by a QSO spectrum that is quite similar in luminosity, radio
core-dominance, and in optical PC1 properties.
We chose 3C\,263\footnote{For PKS\,1004+13, $M_{\rm B}=-25.6$, and
core-dominance = -1.7 (\S4).  Optical PC1 properties are: FWHM(H$\beta$),
6300 km s$^{-1}$; EW ([O\,III]),  6\AA;  [O\,III]/H$\beta$\,intensity ratio, 0.15; and
H$\beta$\ asymmetry, 0.06 (Boroson \& Green 1992).  For 3C\,263, $M_{\rm B}=-27.0$; radio
core-dominance, -1.1; FWHM(H$\beta$), 6100 kms$^{-1}$; EW ([O\,III]), 17\AA;
[O\,III]/H$\beta$\,intensity ratio, 0.24; H$\beta$\ asymmetry $-$0.1.},
with EW(Ly$\alpha$) = 114\AA, and EW(C\,IV) = 86\AA, from
a sample of radio-loud QSOs (Wills et al. 1995).  Figure 3 shows
the result, and serves to illustrate the approximate strength of the
N\,V absorption.  This technique would give an accurate absorption profile only if
the template spectrum were an accurate match to the emission line spectrum of PKS 1004+13,
and in the unlikely situation that the absorbing gas covered the continuum and the
whole of the emission line region.
Note that if the broad absorption in PKS 1004+13 had been further blueshifted,
as in many well-known BAL QSOs, it would have been more clearly seen against a
well-defined continuum.

\section{Discussion}

PKS 1004+13 shows the following characteristics typical of QSOs with high-ionization BALs:
(i) absorption to high outflow velocities ($\sim$10\,000 km s$^{-1}$) in high-ionization
lines of O\,VI, N\,V, C\,IV, and possibly Si\,IV, with a weak Ly$\alpha$\ broad absorption,
and suppression of the broad Ly$\alpha$\ emission line,
(ii) judging by the region of the N\,V `trough' in the high-resolution GHRS spectrum, the troughs
are smooth; probably the C\,IV seen only in low-resolution IUE spectra will not resolve out
in higher resolution observations,
(iii) we measure a non-zero balnicity index, 850 km s$^{-1}$ (cf. Weymann et al. 1991),
(iv) there are multiple (2), high-ionization associated absorption systems, as commonly go along
with BAL absorption (e.g., Ogle et al. 1999),
(v) it is extremely weak in soft X-rays, as in other BAL QSOs (Green et al. 1995; Green \& Mathur
1996),
(vi) there is scattered-light polarization (see below), and
(vii) there is no sign of continuum reddening (Figure 7 of Elvis \& Fabbiano 1984).

Unlike any BAL QSOs reported in the literature\footnote{
Recently, another FR II BAL QSO has been found in the FIRST survey (Becker, private communication).},
PKS\,1004+13 shows classical lobe-dominant radio 
structure (Kellermann et al. 1994,
Miley \& Hartsuijker 1978), with core dominance, log (core flux/lobe flux) = $-$1.74 (5GHz, rest frame).
Thus, for the first time, it may be possible to relate BAL QSO properties to the axis of the central
engine.  To put the core-dominance in context we note that PKS\,1004+13 is also a 4C radio source,
selected at a low frequency, 178MHz.  We can compare the core-dominance with the distribution for
3CR sources, also selected at 178MHz, as this quantity is distributed similarly for all FR\,II
radio sources in 3CR and 4C (Hoekstra, Barthel, \& Hes 1997).  Of 28 3CR QSOs with sufficient radio
imaging, only two have clearly smaller core-dominance -- 3CR\,68.1, an extremely reddened
and highly scattering-polarized QSO with strong associated absorption (Brotherton et al. 1998b), and
3CR\,351 -- with strong absorption by ionized gas in the X-rays and strong, high-ionization, associated
absorption lines in the UV (Mathur et al. 1994).  Of 66 3CR radiogalaxies, half have smaller
core-dominance.  Thus,
in unified schemes for radio sources, this indicates for PKS 1004+13 an inclination angle skimming
the dusty torus, $\sim 50^{\circ}$.
Thus, the BAL region in PKS 1004+13 appears to lie at a large angle to the central engine's
axis, and the BALs could arise in an equatorial accretion disk wind.
This also ties in with pictures of the polarization scattering geometry in which BAL gas lies
at low disk latitudes.

About a third of BAL QSOs show polarization at the $\gtrsim 2$\% level, but this is rare in UV-selected
non-BAL QSOs.  Polarization increases towards the blue as expected for scattered light (Schmidt \& Hines
1999).
PKS\,1004+13 shows continuum polarization at the $\sim 2$\% level, increasing towards the blue
(Antonucci et al. 1996; Stockman, Moore, \& Angel 1984).
The scattered-light geometry suggested by this polarization may be related to the
partial covering of the continuum that we deduce for the O\,VI absorbing region; partial
covering has been
shown directly by changes in polarization across the absorption troughs of some radio-quiet 
BAL QSOs (Schmidt \& Hines 1999).

PKS 1004+13 could have provided an interesting test of scattering geometry for BAL QSOs, through a
comparison of the
polarization position angle with the position angle of the radio axis (117$^{\circ}$).
Unfortunately, the optical position angle is ill-defined at present.
Antonucci et al. and Stockman et al. both find different and
wavelength-dependent polarization position angles, suggesting variability, and probably more than
one polarization mechanism.

Is the discovery of BALs in PKS 1004+13 consistent with the 11\% rate of occurrence
among radio-quiet QSOs (Weymann 1997) or among radio-loud QSOs discovered in deep
radio surveys?  We can address this question by noting that Brandt et al. (1999) have searched
for associated absorption lines in all available HST and IUE spectra with good
signal-to-noise ratio, for all $z < 0.5$\ QSOs in the
PG optically-selected sample.  PKS 1004+130 is the only one of 12 radio-loud
QSOs to show probable BALs -- consistent with the usual discovery rates.
Are BAL properties a function of radio-loudness?  If strong absorption near
$z_{em}$\ is a common characteristic, this may explain why more radio-loud BAL QSOs have not been
discovered.  

PKS\,1004$+$13 is very bright in radio through UV wavelengths, and it is nearby, enabling more
detailed spectroscopic follow-up and imaging at high spatial resolution.  The low redshift also
means that the broad hydrogen lines and narrow line region (NLR) emission lines are accessible
in the optical and near infrared.  High-quality UV spectroscopy is needed to establish whether
PKS\,1004+13 is indeed a BAL QSO, and would, in any case, lead to important clues to physics
and covering factor of the absorbing outflows, the BLR, and the continuum that ionizes it.
Imaging of the relatively bright radio nucleus, polarized (scattered) light and [O\,III]
emission may lead to further clues to the geometry.  
PKS\,1004+13 may therefore be a key AGN in understanding the relationship between radio-loudness 
and BALs, and the geometry of BAL QSO outflows.

\acknowledgments

We thank Michael Brotherton for valuable discussions, and
Nahum Arav for supplying the data for PG\,0946+301.
Thanks to Yoji Kondo and Willem Wamsteker for help in arranging the IUE observations,
and C. Imhoff and R. Thompson for patient answering of IUE questions.
Some data are from the Multimission Archive at the Space Telescope Science Institute (MAST).
STScI is operated by the Association of Universities for Research in Astronomy, Inc., under NASA
contract NAS5-26555.  MAST for non-HST data is supported by the
NASA Office of Space Science via grant NAG5-7584 and others.
This research has made use of the NASA/IPAC Extragalactic Database (NED), which is operated by
the Jet Propulsion Laboratory, Caltech, under contract with NASA.
This research is supported
by NASA through LTSA grant numbers NAG5-3431 (B.J.W.) and NAG5-8107 (W.N.B.).

\clearpage

\begin{figure}
\plotone{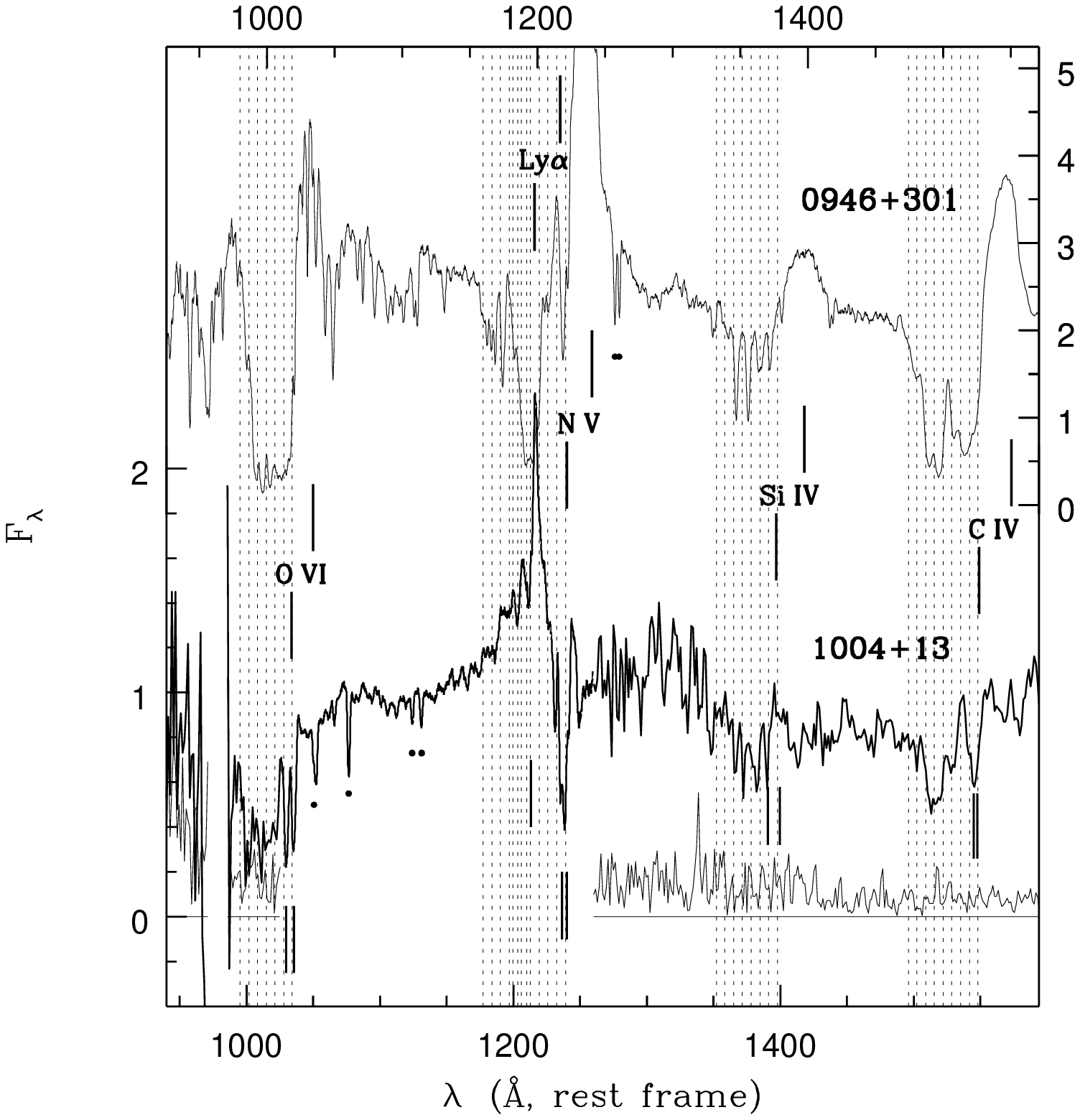}
\caption{Below: The observed ultraviolet spectrum of PKS\,1004+13, derived from HST GHRS and IUE SWP
data, shown on a rest-wavelength scale.  The GHRS data have been smoothed (compare with Fig.\,2).
The standard deviation spectrum is shown, only for the IUE data.
Above: The smoothed HST spectrum of the well-known BAL QSO, PG\,0946+301
(Korista \& Arav 1997), on a redshifted scale (top) that aligns its BALs with our suggested BALs for 
PKS\,1004+13.  The ordinates are in 10$^{-14}$\,erg s$^{-1}$\,cm$^{-2}$ \AA$^{-1}$\ and 
10$^{-15}$\,erg s$^{-1}$\,cm$^{-2}$ \AA$^{-1}$\ for PKS 1004+13 and PG 0946+301, respectively.
The BAL velocity range, 0 to $-$10,000 km s$^{-1}$, is indicated by the
shading.  Note the overlapping troughs of Ly\,$\alpha$\ and N\,V\,$\lambda$1240.  The expected
wavelengths of emission-line peaks are indicated next to the labels (for PKS\,1004+13, 
$z_{\rm em} = 0.2401$; for PG\,0946+301, $z_{\rm em} = 1.222$).  The expected positions of narrower
absorption doublets are shown below the PKS\,1004+13 data.  Dots show the positions of Milky Way
interstellar absorptions.  The gap near 980\AA\ corresponds to geocoronal Ly$\alpha$.
}
\end{figure}

\clearpage

\begin{figure}
\plotfiddle{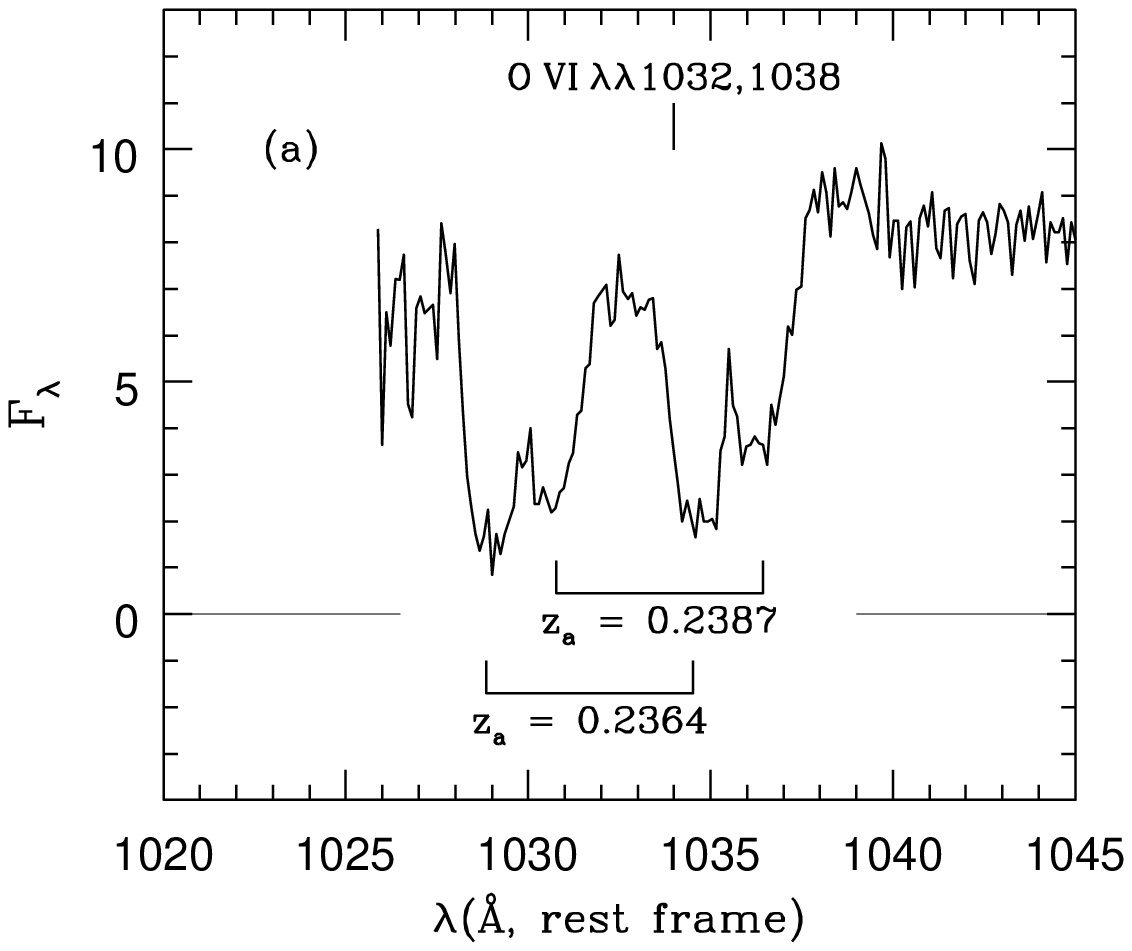}{5.0cm}{0.}{50.0}{50.0}{-100.}{-80.}
\plotfiddle{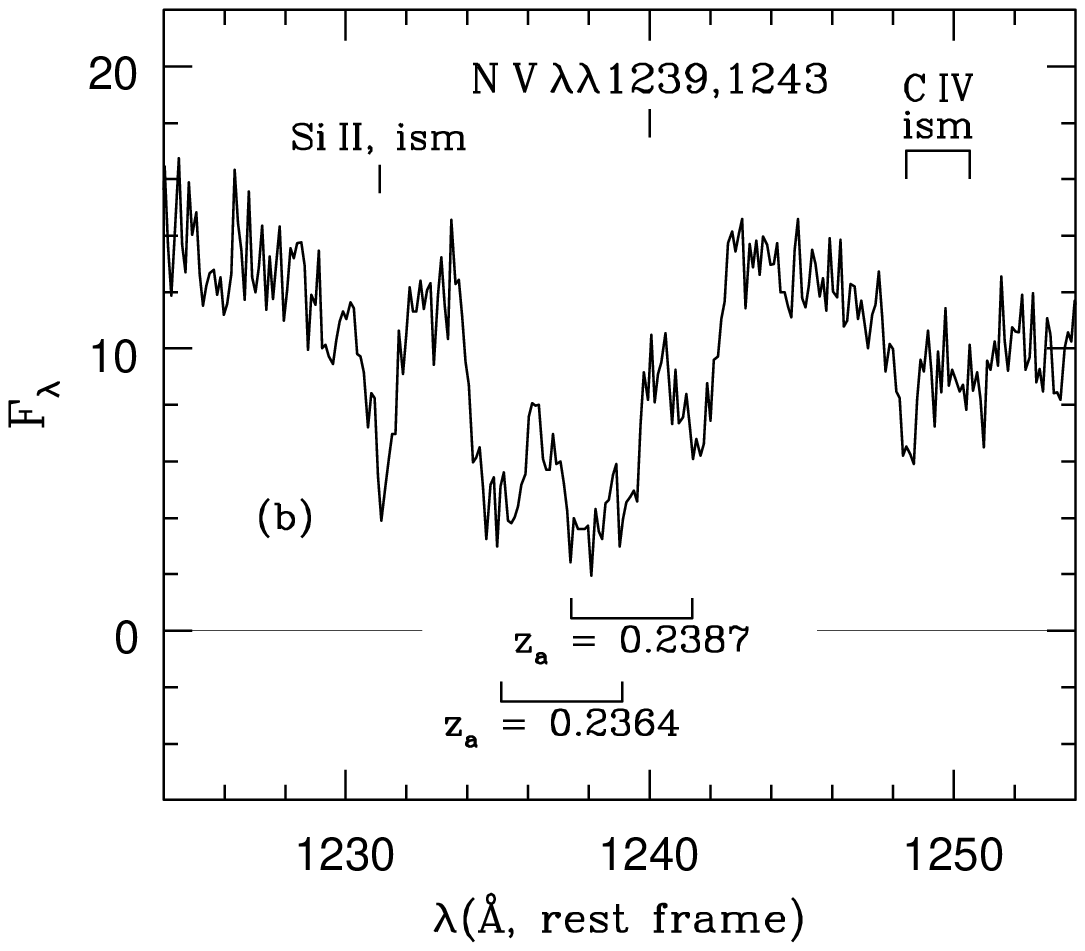}{5.0cm}{0.}{50.0}{50.0}{-100.}{-80.}
\plotfiddle{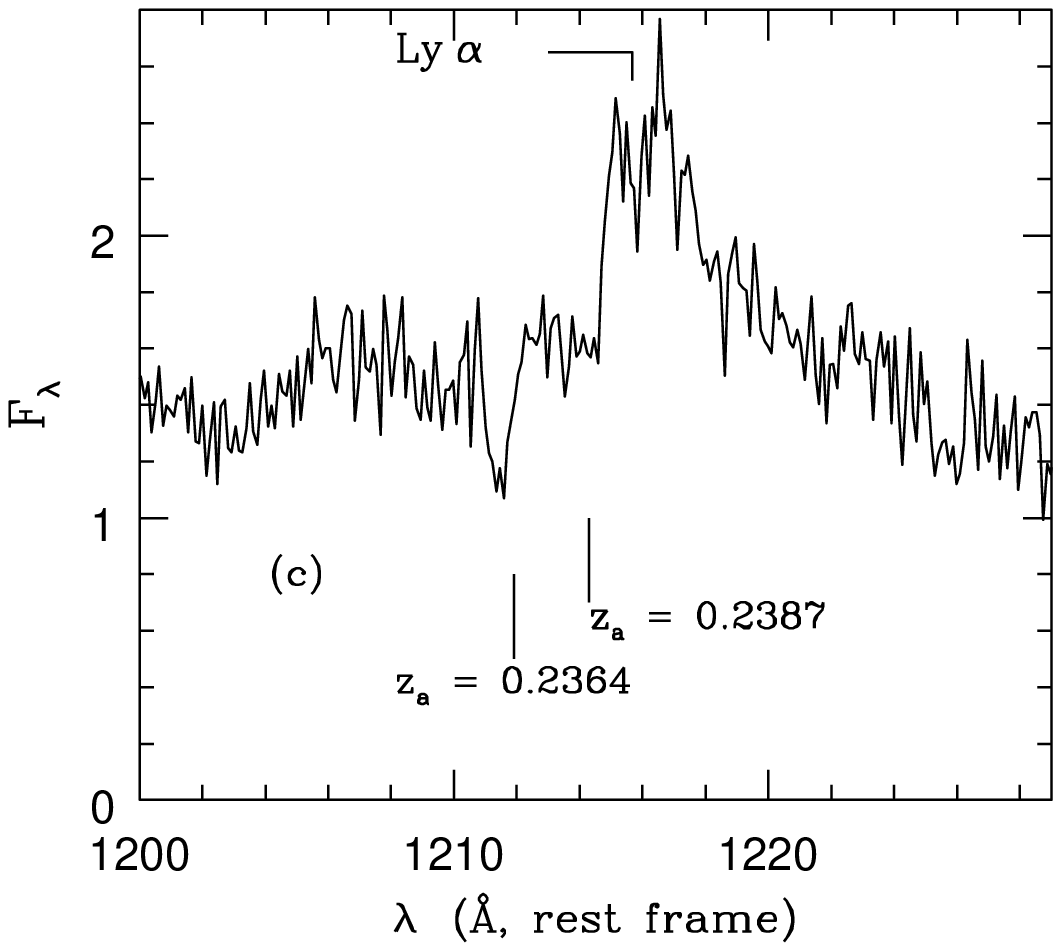}{5.0cm}{0.}{50.0}{50.0}{-100.}{-80.}
\caption{The narrower absorption features in PKS\,1004+13, illustrating the two
redshift systems, and the widths of the absorption features for the wavelength
regions of a) O\,VI, b) N\,V, and c) Ly$\alpha$.  The expected wavelength of the emission-line
peak is indicated above the spectra.  The long-wavelength edge of the absorption corresponds to 
zero rest-frame velocity for the longest wavelength pair of the doublets. 
For the ordinate scale, see Fig.\,1.
}
\end{figure}

\clearpage

\begin{figure}
\plotone{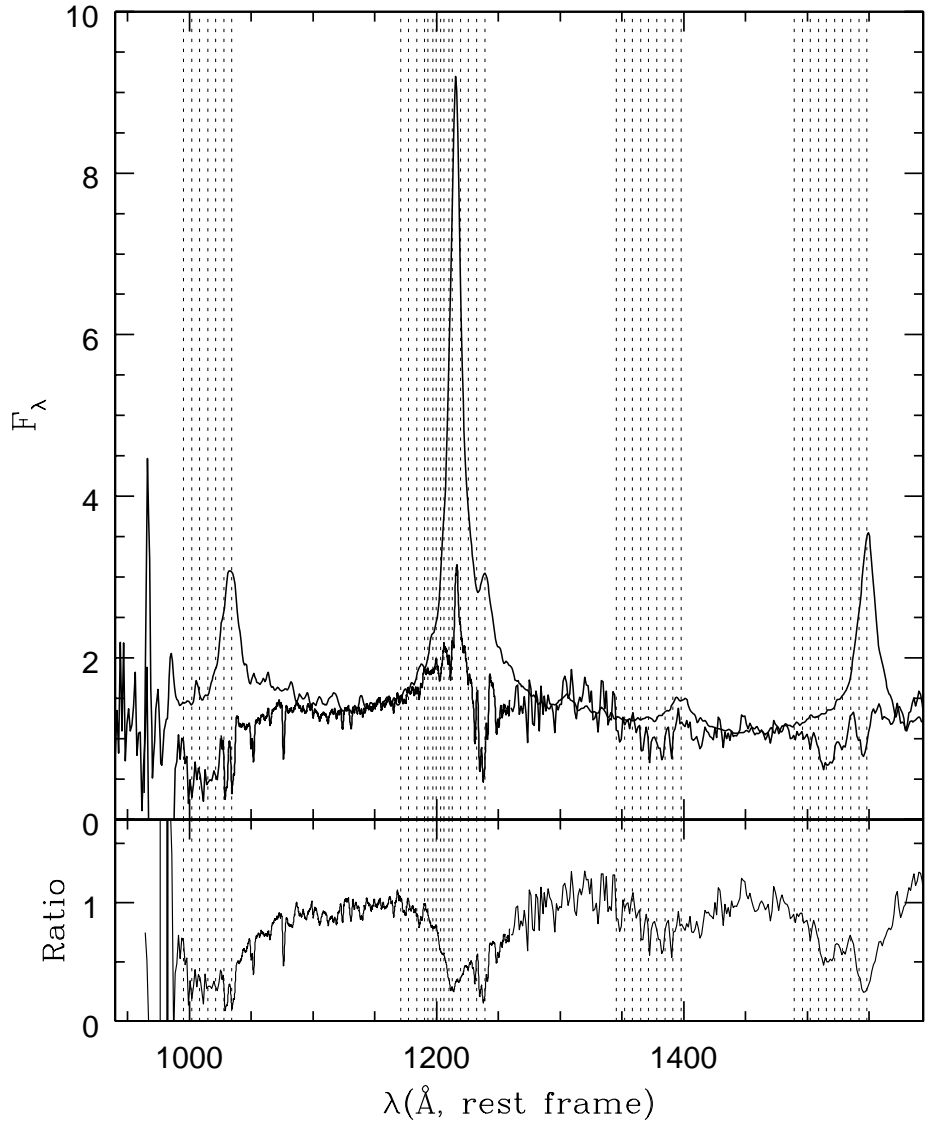}
\caption{
Comparison of a smoothed, scaled spectrum of a matched radio-loud QSO 3C\,263 (see text) 
with the ultraviolet spectrum of PKS\,1004+13.  Both are corrected for Milky Way reddening.
In the lower panel we show the division of the spectra, PKS\,1004+13 by 3C\,263.
}
\end{figure}

\end{document}